\documentclass[sigconf]{acmart}

\usepackage{hyperref}           
\hypersetup{
    colorlinks=true,
    linkcolor=blue,
    filecolor=red,      
    urlcolor=magenta,
    breaklinks=true,            
}
\usepackage{breakurl}           
\usepackage{amsmath,amsthm,xcolor}

\newtheorem{defn}{Definition}
\newtheorem{axiom}{Axiom}

\AtBeginDocument{%
  \providecommand\BibTeX{{%
    \normalfont B\kern-0.5em{\scshape i\kern-0.25em b}\kern-0.8em\TeX}}}

\copyrightyear{2021} 
\acmYear{2021} 
\setcopyright{acmlicensed}
\acmConference[FAccT '21]{Conference on Fairness, Accountability, and Transparency}{March 3--10, 2021}{Virtual Event, Canada}
\acmBooktitle{Conference on Fairness, Accountability, and Transparency (FAccT '21), March 3--10, 2021, Virtual Event, Canada}
\acmPrice{15.00}
\acmDOI{10.1145/3442188.3445943}
\acmISBN{978-1-4503-8309-7/21/03}

\begin{document}

\title[Epistemic values in feature importance methods]{Epistemic values in feature importance methods: Lessons from feminist epistemology}


\author{Leif Hancox-Li}
\affiliation{%
  \institution{Capital One}
  \streetaddress{11 W 19th St}
  \city{New York, New York}
  \country{USA}}
\email{leif.hancox-li@capitalone.com}

\author{I. Elizabeth Kumar}
\affiliation{%
  \institution{University of Utah}
  \city{Salt Lake City, UT}
  \country{USA}
}
\email{kumari@cs.utah.edu}


\begin{abstract}
As the public seeks greater accountability and transparency from machine learning algorithms, the research literature on methods to explain algorithms and their outputs has rapidly expanded. Feature importance methods form a popular class of explanation methods. In this paper, we apply the lens of feminist epistemology to recent feature importance research. We investigate what epistemic values are implicitly embedded in feature importance methods and how or whether they are in conflict with feminist epistemology. We offer some suggestions on how to conduct research on explanations that respects feminist epistemic values, taking into account the importance of social context, the epistemic privileges of subjugated knowers, and adopting more interactional ways of knowing.

\end{abstract}


\keywords{explanation, philosophy, epistemology, machine learning, feature importance, feminism, methodology}

\maketitle

\section{Introduction}

In recent years, the number of new methods for measuring feature importance for machine learning (ML) models has exploded, leaving ML practitioners spoilt for choice. As black-box algorithms with inscrutable inner mechanisms are increasingly used for crucial decisions, demands for greater transparency and accountability have increased, leading to legal requirements for explanation like that in the European Union's General Data Protection Regulation (GDPR). Media coverage and public awareness of potentially problematic algorithmic decisions have also increased, forcing ML practitioners to come up with improved and novel ways to explain their models. In this paper, we focus on a subclass of explanation methods, known as feature importance or feature attribution methods. Broadly construed, these assign a quantitative measure of importance to the inputs to a function. They are a popular way of approaching demands for explainability, as they fit the human tendency to attribute outcomes to specific factors.

As feature importance methods proliferate, practitioners struggle to decide which methods to use, and creators of methods strive to demonstrate that their methods are superior to others. However, much of the evaluation of various methods' pros and cons proceeds in an implicitly value-neutral context: rarely do participants in this discourse pause to consider if certain supposed theoretical or practical virtues would be virtues in all social or applicational contexts.\footnote{The exceptions to this tend to occur in interdisciplinary venues like the FAccT conference \cite{barocas-counterfactual}, or in venues for human-computer interaction professionals \cite{kaur, hong2020human}.} Papers introducing new feature importance methods state that certain properties of the proposed methods are ``desirable'' without indicating the contexts or uses for which those properties are desirable \cite{ig, shap}. Stating a property as ``desirable'' without further elaboration elides questions such as: Desirable to whom? For what purposes? When applied to what types of data?

In this paper, we argue that a number of popular feature importance methods implicitly contain values that are counter to the kind of pluralistic, contextual, and interactional view of epistemology advocated by many feminist epistemologists.
The existence of these values is not problematic in itself unless one adheres to the ideal of a ``view from nowhere'' \cite{nagel1989view}. However, the partiality of feature importance methods is often not acknowledged by the creators or users of those methods. Instead, debates focus on various apparently value-neutral properties of the methods, to the point where some aspire to create universal benchmarks to evaluate all explanation methods \cite{doshivelez2017rigorous, schmidt2019quantifying}.

In the next section, we provide a brief overview of feminist epistemology, in particular, the frameworks of situated knowledges and standpoint epistemology. In Section \ref{values}, we describe the various ways in which values counter to feminist epistemology may be embedded in feature importance methods. Finally, in Section \ref{constructive}, we end by offering some suggestions on how ML research on explanations can include more feminist values.

\section{Feminist Epistemic Values}
\label{feminism}

Against the unitary view from nowhere that is assumed in much of science and mainstream analytic epistemology, feminist epistemologists have argued for understanding science through the idea of situated knowledges. Instead of aiming for one correct, objective view of things, they argue that we should instead accept that knowers are social beings, who bring perspectives to each issue that are conditioned by their social experiences. In this view, there is no ``god trick'' that lets us know from the point of view of the unmarked \cite{haraway1988situated}. It is not, however, a completely relativistic view---we can still speak of views being ``less false'', even if we cannot speak of truth or truth-likeness \cite{harding1995strong}.

Feminist epistemologists came to this view as a reaction against what they saw as problematic claims to unitary knowledge in professions, like the natural sciences, where largely white and male knowers have constructed one way of seeing as the ``objective'' one, while alternative ways of knowing, many propounded by marginalized groups in science, are ignored \cite{keller1995reflections}. Feminist epistemologists have described numerous examples of how the natural sciences could have benefited from accommodating more ways of knowing \cite{keller1984feeling, martin1991egg}. Indeed, some argue that subjugated knowledges, that is, the perspectives of those with the least power, have an advantage in providing more empirically accurate and comprehensive accounts of the world \cite{harding1995strong, Wylie2015}. Given the non-representative nature of machine learning practitioners' demographics \cite{ainow2017}, we should be similarly wary of claims by ML practitioners about certain methods being universally or objectively better than others.

Indeed, compared to many of the natural sciences examined by feminist epistemologists, there may be even greater reason to be wary of claims of value-neutrality in ML, because much of the value of ML explanations comes from their utility in practical contexts. The same type of explanation may be useful in one type of practical application but not another.\footnote{For example, a feature-highlighting explanation may provide an actionable guide to a decision subject on how to change an algorithmic decision in some contexts, but not in others, depending on what factors the explanation recommends that the subject change \cite{barocas-counterfactual}. Some factors are impossible to change due to physical impossibility or environmental factors. However, the same feature-highlighting explanation that is unhelpful to a decision subject may be useful to a data scientist doing feature selection or feature engineering.} This is unlike many hypotheses in the natural sciences, which are often evaluated not solely by  their practical utility, but also by how they fit with existing scientific theory as part of a coherent, realistic description of the world.

In contrast to the view from nowhere, the situated knowledge framework focuses on the pragmatics of acquiring knowledge in particular contexts. This kind of contextual thinking includes considering how traditional epistemic goals intersect with social goals. As Sandra Harding points out, ``The kinds of explanations favored by modern science have not always been the most effective ones for all projects---for example, for achieving environmental balance or preventing chronic bodily malfunctions'' \cite{harding1995strong}. In the case of ML explanations, the intertwining of epistemic and social goals is particularly striking, since the demand for an explanation is often not just a demand for knowledge of how the algorithm works, but also part of an ethical demand for greater accountability or transparency \cite{doshi2017accountability, rader-transparency}.

Another aspect of feminist epistemology that is particularly relevant to ML explanations is its implications for formal frameworks. As Donna Haraway argues, ``rational knowledge does not pretend to disengagement... to by [sic] fully self-contained or fully formalizable. Rational knowledge is a process of ongoing critical interpretation among `fields' of interpreters and decoders'' \cite{haraway1988situated}. In reducing a complex system to one formal framework, we often reduce the possible meanings it can have and the possible relations it can have to other systems. Focusing on formalisms also distracts from the meanings we attach to formal symbols or formulae, even though the same formalism may take on different meanings in different social or applicational contexts.

Haraway also advocates for a critical stance on boundaries and objects. By paying attention to how objects of study are partly constructed by social processes, we can also see how the boundaries that define those objects ``materialize in social interaction'' \cite{haraway1988situated}. Similarly, in taking a feminist approach to ML explanations, we should critically question how we draw the boundaries around what the realm of ``explanation'' is. Are we defining the realm of legitimate study of explanations in an unnecessarily narrow way? Are there possible agents who can help construct explanations or be part of explanations, that we have left out of our idealized view of explanations?

Finally, feminist epistemology also encourages a more interactive way of how knowledge is created,  considering the object of knowledge not just as a passive object with static properties, but also as an agent that can enter in conversation with the knower. In this view, ``[t]he world neither speaks itself nor disappears in favor of a master decoder'' \cite{haraway1988situated}. For example, Evelyn Fox Keller attributes Barbara McClintock's successful insight into maize genetics to McClintock's willingness to erode the boundaries between subject (the scientist) and object (the maize plants) \cite{keller1995reflections}. Keller also attributes McClintock's openness to questioning the central dogma of molecular biology to her lack of investment in the idea of the passivity of nature. A related idea, from black feminist epistemology, is that of using dialogue to assess knowledge claims---interacting with the object of your knowledge rather than observing it from a detached distance \cite{collins1990black}.

In short, feminist epistemology provides a more pluralistic, contextual, interactional, and critical corrective to various traditionally ``masculine'' ways of doing science, a corrective that can also be applied to how we create, use, and evaluate feature importance methods in ML. As we'll see in the next section, the apparently objective endeavor of figuring out how to attribute feature importances contains various implicit values, many of which are counter to the principles of feminist epistemology.

\section{Values in Feature Importance Methods}
\label{values}

We define feature importance to be any quantitative assignment of importance or influence to each input feature of some model learned from data. How the notion of ``importance" is formalized varies between methods, and can be defined in reference to either the entire model and its training procedure, or to the model's prediction on one particular input.

To provide broader context for our more general claim that feature importance methods can be inflected with particular epistemic values, a look at the history of feature importance methods is instructive. This history shows how contingent choices influenced by practitioners' social environments led to the current state of affairs.  In particular, the widespread modern emphasis on predictive accuracy as the preeminent virtue of feature importance comes from instrumentalist epistemic values championed by the likes of Leo Breiman.

Informal descriptions of the utility of feature importance often cite the high dimensionality of useful models, implying that they could be made smaller \cite{lam1999feature,7879832,guyon2003introduction}. In the popular textbook ``Elements of Statistical Learning,'' for instance, Hastie, Tibshirani and Friedman write, ``In data mining applications the input predictor variables are seldom equally relevant. Often only a few of them have substantial influence on the response; the vast majority are irrelevant and could just as well have not been included. It is often useful to learn the relative importance or contribution of each input variable in predicting the response'' \cite{hastie2009elements}. Indeed, in the traditional, statistical data-modeling setting, feature importance is often described in terms of feature selection (such as in the case of F-tests), with the ultimate goal being a simplified model that is more likely to describe a real phenomenon. Data modeling in this sense has an emphasis on description, that is, the objective of scientific truth.

In describing a change of direction away from data modeling, Matthew Jones argues that modern research in data science is born ``more from an engineering culture of predictive utility than from a scientific culture of truth" \cite{jones2018we}, and the same can be said of feature importance. This can be traced to Leo Breiman's novel proposal of using feature importance for purposes other than feature selection. Breiman argued against ``assuming that the data are generated by a given stochastic data model'' in favor of a new algorithmic culture \cite{breiman2001}.

Breiman proposed a new, influential view of data analysis which de-emphasized the importance of developing probabilistic models of underlying data generating processes. He argued that a prediction-centered view was the right way to investigate the relationship between the response and predictor variables, and that black-box algorithmic models can provide more reliable and interesting information than weakly predictive data models can: ``The goal is not interpretability, but accurate information. Interpretability is a secondary goal that can be finessed.'' Regardless, he proposed a general way to measure the importance of predictors in a model: ``My definition of variable importance is based on prediction. A variable might be considered important if deleting it seriously affects prediction accuracy'' \cite{breiman2001}.

This logic makes sense on its face, but Breiman never worked towards providing a ``satisfactory theoretical definition'' \cite{breiman2001} of importance. The mechanism of variable deletion, a step in the process of feature selection, was the basis of the analysis; but for Breiman, the advantage of using trees and ensembles of trees was exactly that it allowed one to avoid doing feature selection: ``Existing methods, he and his collaborator Meisel argued in a report to the US Air Force, required the analyst to choose, without reference to the data, a means for reducing its dimensionality'' \cite{jones-forthcoming}. Breiman's version of feature importance was not intended to help create more veridical models of reality by deleting extraneous information, and this is precisely the reason that permutation feature importance, his approximation of the definition he intuited, has been shown to be problematic \cite{hooker2019please}. Breiman wanted to calculate the change in prediction accuracy under the removal of information, but the removal is simulated instead of actually tested because there was no intention of actually removing the feature.

Yet permute-and-predict notions of importance persisted throughout recent algorithmic history. Hooker and Mentch observed that this was because of their pragmatic properties: ``they are each computationally cheap, requiring $O(N)$ operations, apply to the $f(x)$ derived from any learning method, have no tuning parameters, and in relying only on averages, they are statistically very stable. Moreover, the approach is readily understood and easily explained to users in applied areas'' \cite{hooker2019please}. Permutation feature importance influenced many ideas throughout the history of the problem of calculating feature importance for black-box models  \cite{altmann2010permutation, mcr, covert2020understanding, auditing}. As argued in \cite{dotan-milli}, accuracy is not necessarily a value-neutral evaluation criterion; the popularity of black-box algorithmic models reflect a value-laden disciplinary shift in the field of machine learning, and this is paralleled in the history of feature importance.

While instrumentalism of the kind advocated by Breiman may not be particularly feminist or anti-feminist, it's certainly an epistemic value that became dominant due to historically contingent factors. Replies by Brad Efron and Bruce Hoadley in \cite{breiman2001} argued against Breiman's conception  of feature importance and indicate a possible alternate path for feature importance that failed to become the dominant conception. Absent the social factors that influenced Breiman's thinking and that facilitated his stature in the discipline, perhaps we would have discussions of feature importance that have a more pluralistic tone, rather than discussions that over-emphasize the same ``virtues''. The anti-pluralism inherent in Breiman's championing of predictive accuracy can be seen as counter to feminist epistemology. In addition, some feminists have regarded instrumentalist values as counter to feminism.\footnote{See \cite{hirschheim1996exploring} for an example of this view, and, as a counterpoint, \cite{adam2001feminist} for why anti-positivism or anti-instrumentalism may be unhelpful for feminist goals.}

This historical preamble provides crucial context for what follows, by demonstrating that the epistemic values favored today are the product of historical processes influenced by social factors. In the following sections, we describe more epistemic values that inflect modern feature importance methods, arguing that these values conflict with feminist epistemology.

\subsection{Universality as an epistemic value}
\label{universal}

One aspect of how values enter into feature importance measures is the ways in which these are derived and evaluated. Many methods are arrived at axiomatically, and many proposed evaluation standards pin their hopes on a universal ``benchmark'' or ``gold standard'' \cite{doshivelez2017rigorous, schmidt2019quantifying}. We argue that the axiomatic approach towards derivation and the ``gold standard'' narrative for evaluation embed universal, anti-pluralist values, which go against feminist epistemic values \cite{adam2001feminist, d2020data}. At the same time, human-centered studies on how explanation methods are used in real-world contexts demonstrate that these methods are used for very different purposes and audiences, cautioning against applying context-free desiderata or evaluation criteria \cite{hong2020human, hohman2019gamut}.

Many proposals of new explanation techniques take an axiomatic approach, where they specify desirable properties for explanations, then construct or derive a method that has those properties \cite{shap, ig, sliwinski2019axiomatic, pmlr-v97-chattopadhyay19a, singal2019, bhatt2020evaluating}. They also often cite the fact that a method ``uniquely'' satisfies some set of axioms as a reason to accept it, such as in local explanations based on Shapley values \cite{shap, st_interaction}. The Shapley value is the solution to a fundamental problem in game theory, which is uniquely defined with respect to certain axiomatic conditions put forth by many authors as fundamentally desirable. However, even in the original context of game theory, some of these axioms were thought by mathematicians to be ``mathematically convenient" \cite{osborne1994course} and ``not so innocent" \cite{luce1957games}. By the former, they meant that the solution is constrained to be unique only if it is required to satisfy all of the axioms, which makes it tempting to say that the axioms are desirable. Despite this, the creators of SHAP, a popular feature importance method based on Shapley values, explicitly position their method as embodying what they hope will be universally desirable principles for all explanation methods:
``The thread of unity that SHAP weaves through the literature is an encouraging sign that common principles about model interpretation can inform the development of future methods''  \cite{shap}.
Similar remarks on unity can be found in further elaborations on SHAP \cite{covert2020understanding}.

There are some understandable reasons for taking an axiomatic approach. As \cite{ig} argues, when evaluating an explanation method, it can be unclear if a provided explanation looks wrong because the model itself is wrong, or if it looks wrong because the explanation method is at fault. The axiomatic approach sidesteps these empirical ambiguities by providing a guarantee that the method has certain ostensibly desirable mathematical properties, regardless of its empirical performance. However, proponents of axiomatic approaches position their methods as providing ``intuitively'' desirable properties \cite{covert2020understanding, ig}. Implicit in this way of justifying axiomatic properties is that the intuition referred to is universally held or applicable to all situations.\footnote{See below for examples of specific axioms that are unlikely to be intuitive or valid in many real-world situations.} This ignores the fact that those properties may not be the most appropriate ones in every context of application.

A more pluralistic view of explanatory virtues would cast a critical eye on the idea of universally desirable properties and the ability of a researcher to know what those are \emph{a priori}. User studies back up the idea that explanation methods are used in very different contexts, for different purposes \cite{hong2020human, hohman2019gamut}. In the absence of user studies comparing the usefulness of explanation methods for different purposes, it's not obvious that a method that is more effective for communicating with non-technical stakeholders is necessarily also more effective for helping data scientists debug a model. As a concrete example, one user study found that SHAP is not very helpful for helping data scientists spot obvious problems with the data \cite{kaur}, but another user study found that SHAP is good at producing explanations that fit human intuition \cite{lundberg2018consistent}. As we point out in Section \ref{context}, many papers on ML feature importance methods evaluate the methods on ``tasks'' that are untied to any specific pragmatic goal, even as the feminist epistemology literature has long pointed out the problems with ``supposedly universal principles drawn from bounded artificial examples" \cite{adam2001feminist}. 

In the broader context of ML research in general, others have warned of the pitfalls of having benchmarks that determine what ``success'' or ``progress'' is \cite{dotan-milli,denton2020bringing}. Even the organizers of ML competitions admit that competing on fixed metrics can lead to a kind of meaningless optimization after a while \cite{everingham2015pascal}. These broader lessons also apply to ML explanations. Favoring a fixed number of benchmarks for ML explanations risks channelizing new methods into a few directions, privileging some goals or audiences over others. Without broader participation from different stakeholders in the design of ML explanations, this runs the risk of not just restricting directions of research, but also creating explanation methods that privilege the interests of certain groups over others.

\subsubsection{Additively partitioning influence}

We've argued that the idea of a universally applicable explanation method, based on universally desirable axioms, is contrary to pluralism. In addition, some axioms that are touted as obviously desirable contain their own embedded values. One type of axiom that is common in axiomatic approaches towards explanations is an axiom that encodes the additivity of feature importances.

The idea behind this notion is that the individual contributions of each feature, as measured by the feature importance method, should add up to the interventional effect of all the features taken together, thus partitioning the total effect of the feature set among its components. This axiom either goes unnamed \cite{lrp} or appears under different names in different papers, such as ``completeness'' \cite{ig}, ``summation-to-delta'' \cite{shrikumar2017learning}, or  ``local accuracy'' \cite{shap}.  In particular, given a machine learning model $f$ which computes a prediction for a specific instance $x$, the axioms can be described as the following.

\begin{axiom}[Completeness]
Given feature attributions $\phi_1, \phi_2, ... \phi_n$ corresponding to features 1 through $n$ for a certain input $x$ to a function $f$, $f(x) = f(x') + \sum_{i=1}^n \phi_i$ for some baseline input $x'$.
\end{axiom}

\begin{axiom}[Local Accuracy]
Given feature attributions $\phi_1, \phi_2, ... \phi_n$ corresponding to features 1 through $n$ for a certain input $x$ to a function $f$ and its simplified features $h_i(x)$ representing the ``presence" of feature $i$, $f(x) = \phi_0 + \sum_{i=1}^n \phi_i h_i(x)$ for some baseline $\phi_0$.
\end{axiom}

These two axioms are equivalent if we take $\phi_0 = f(x')$ and define $h_i(x)$ to be 1 for all $i$. 

A stronger desideratum, which requires the sums of individual contributions to be proportional to the joint effect for each \emph{subset} of features, is called faithfulness \cite{bhatt2020evaluating}. Faithfulness is proposed as a metric, rather than an axiom, which measures closeness to an ideal: a perfect faithfulness score implies completeness and local accuracy.

\begin{defn}[Faithfulness]
Given feature attributions $\phi_1, \phi_2, ... \phi_n$ corresponding to features 1 through $n$ for a certain input $x$ to a function $f$, for any subset $S$, let $x_S$ be equal to $x$ with the features not in $S$ replaced with those from some baseline vector $x'$. Then faithfulness is defined as $$\text{corr}_{S \subseteq \{n\}} \left( \sum_{i \in S} \phi_i, f(x) - f(x_S) \right)$$
\end{defn}

Axioms like completeness, local accuracy, and faithfulness align with the natural properties of coefficient size in linear models of independently distributed features. \footnote{The related axiom of linearity in \cite{ig}, that linear combinations of models should result in linear combinations of feature attributions, is justified ``intuitively." Linearity is implied by how we would want coefficients to combine if we were talking about linear regression. This ``intuition" thus again presumes that coefficient size is a reasonable way to explain linear models.}
Given a linear model $f(x) = \beta_0 + \sum_{i=1}^n \beta_i x_i$, it is easy to see what happens when we change the features in some set $S$ to be taken from some baseline vector $x'$: the difference between $x$ and $x_S$ is exactly $\sum_{i \in S} \beta_i (x_i - x'_i)$. Thus, the feature importance term $\phi_i = \beta_i(x_i - x'_i)$, where $x'$ is taken to be the empirical mean of the data, satisfies Completeness and Local Accuracy, as well as minimizing Faithfulness, and is in fact the value assigned by SHAP for this model. In this linear setting, $\phi_i$ represents the univariate effect of perturbing feature $i$ from its expectation to the value represented in the input.

If a user interprets feature importance to represent univariate effects in \emph{all} modeling settings, however, this property is fundamentally misleading in many applications of complex ML models. Often, input features are highly correlated, and ML models learn to infer complex interactions between them. Additivity axioms seek to impose an ideal which does not hold in the real world. Imposing additivity in real-world conditions thus attempts to optimally simplify real interactions in the data in order to fulfill some ideal of analytic and cognitive tractability. In doing so, researchers privilege achieving some kind of mathematical ideal over modeling the genuine complexity in the data.

This is not to say that it is always bad to impose any of these axioms. Rather, they serve as examples where a value-laden choice is made, namely, to stipulate a mathematical ideal to obtain certain analytic goals as opposed to creating new tools to handle complexity at face value. This \emph{starting} from what is formally tractable means that ``what is deemed interpretable comes to affect what is presented as explainable, and thus what can be considered meaningful in an algorithmic system'' \cite{benjamin2019materializing}. Other potential sources of meaning are foreclosed.

There might be certain purposes for which favoring a mathematical ideal is appropriate. But for other purposes, such as understanding complex interactions between features without over-simplifying, it is less appropriate. The value-laden nature of this design choice in creating explanation methods is hidden behind the apparently objective and ``rigorous" approach of starting from axioms and proving uniqueness theorems based on them.

It is important to acknowledge that the idealized nature of the properties discussed above aren't just a mathematical ideal, but also a cognitive ideal: it is thought that it's easier for people to understand and use explanations if the feature attributions behave as described in the relevant axioms. But again, if the actual features do not behave as stipulated in the axioms, then the cognitive ideal merely gives the \emph{appearance} of understandability. This is potentially dangerous when it leads to data scientists having unjustified confidence in their model and thus to premature intents to deploy, as found in \cite{kaur}. In addition, to the extent that there are cognitive limitations on the kinds of explanations that can be understood, this doesn't mean that the correct way to impose those limitations is by the deductive approach of starting from axioms that translate those limitations into mathematics. While the deductive approach may be more in line with conventions of scholarship in machine learning, perhaps more inductive approaches that \emph{start} with empirical work on what users need in specific contexts, rather than using user experiments to validate a deductive approach after the fact, would lead to different types of explanation methods.

\subsection{Epistemic values in the data-versus-model debate} 
\label{data-vs-model}

Another way in which epistemic values may enter into particular definitions of a feature importance method is the extent to which the explanation emphasizes structure in the data, as opposed to structure in the model. This contrast is pertinent in the ongoing debate about the appropriate type of probability distribution to use in computing various feature importance measures, including (but not limited to) SHAP \cite{kumar2020problems}. The choice is between using ``interventional'' distributions, which ignore correlations between input variables, and using ``observational'' distributions, which take those correlations into account.

Specifically, given an input $x$ to a function $f$ and a specific feature set $S$, the question is how to quantify the effect of removing information about $S$ by looking at the value of the quantity $f(x) - f(x_S)$, where $x_S$ is $x$ but with the features in $S$ replaced by samples from some distribution. One option is to \emph{intervene} on those features, using (for instance) their marginal distributions or a fixed baseline value, as in \cite{datta, merrick2019explanation}; the other is to condition on the features not in $S$ and sample from what the values in $S$ might be from the other features in $S$, based on correlations in the \emph{observational} data.

The authors of SHAP recommend using interventional distributions if we want the explanation to be ``true to the model'', and observational distributions if we want the explanation to be ``true to the data'' \cite{chen2020true}. In short, we should use interventional distributions if we want to know how the model works independently of the structure of the particular data it's used on. On the other hand, they argue, if our interest is in a ``natural mechanism in the world,'' then the explanation derived from observational distributions captures the important causal features more accurately.

Going back to the importance of contextual thinking in feminist values, we can see that the ``true to the model'' option abstracts the model away from its application context, which is particularly unhelpful in assessing its impact on real populations. The authors' justification for interventional distributions in this option is illuminating, as they use an example that is particularly ill-suited for interventional distributions once we carefully consider the context of application.

To compare interventional and observational distributions, the authors of \cite{chen2020true} consider a model that predicts whether an applicant will default on a loan. They then test how useful the interventional and observational SHAP values would be to potential applicants, by considering how effectively an applicant can modify their risk by setting a feature to its mean. However, this way of testing is decontextualized from the reality of the deployment context, because in the real world, it's often impractical for most applicants to be able to change only one feature without also changing others \cite{barocas-counterfactual}. Based on this method of testing, the authors conclude that the interventional distribution is better if we want to be ``true to the model,'' and in addition that ``[b]eing true to the model is the best choice for most applications of explainable AI, where the goal is to explain the model itself.'' This statement reveals the model-centered orientation that is an implicit epistemic value in their research. A countervailing perspective might consider that many real-world applications work on data that's highly correlated, in contexts where individuals are often not able to change the value of one feature at a time.\footnote{In particular, the credit data that the authors use in their experiment is probably highly correlated and therefore hard for decision subjects to intervene on ``cleanly'' in the way the authors do in the experiment. The original dataset used in \cite{chen2020true} has been removed from public view, but given that it's in the context of credit, the criticisms of \cite{barocas-counterfactual} on how feature-highlighting explanations of credit models are hard for decision subjects to act on are probably applicable to it.}  The authors implicitly center understanding the model independently of application context. In contrast, feminist epistemologists often criticize the use of simple examples and advocate considering real-life examples in all their complexity \cite{adam2001feminist}.

Other researchers have taken sides in this debate in the context of their own work. The authors of \cite{ig}, for instance, argue that implementation invariance is a desirable property of feature importance: if two different models practically describe the same function, their feature attributions should be the same. Asserting that the mechanics of how the inputs relate to the outputs are not important reflects the rejection of data modeling as understood in \cite{breiman2001}. The authors of \cite{ig} additionally assert that we should not be able to attribute influence to features not in the model, and instead only explain the model itself. \citeauthor{merrick2019explanation} share this view, calling features not in the model ``irrelevant" \cite{merrick2019explanation}. This mathematically implies the choice of interventional distributions in SHAP, as any influence notion that can be captured through correlation could assign influence to unmodelled variables. Although individual counterarguments to these ideas have been pointed out \cite{hooker2019please, slack2020fool}, they remain a persistent trend in the explainable ML literature.

\subsection{Computational modularity as an epistemic value}

Another way in which post-hoc feature importance methods contain embedded values is in design decisions that abstract from real-world complexities and hide implementation details behind an easy-to-use interface, all in a plug-and-play package. Previous research on values in computer science has identified this as a key part of computational culture \cite{malazita2019infrastructures, golumbia2009cultural}, and indeed, this kind of modular design, where details are hidden except on a need-to-know basis, is part of what constitutes good design in programming \cite{evans2004domain}. Other researchers have cautioned against the same tendency when it comes to ``fairness" solutions in machine learning, calling it the ``Portability Trap'' \cite{selbst2019fairness}. Lucy Suchman has characterized this tendency as ``construing technical systems as commodities that can be stabilized and cut loose from the sites of their production long enough to be exported \emph{en masse} to the sites of their use'' \cite{suchman2002located}. Assuming that a solution is portable in this way is possible only if you think contextual factors (e.g. the circumstances that the model and explanation are used in) are less important than the solution itself.

In the context of feature importance methods, hiding the details is not a value-neutral decision, especially when the very virtues of the methods themselves are under debate. User research has shown that many data scientists interpret SHAP outputs in a relatively uncritical manner. Many are unable to describe the visualizations produced by these methods, even though they trust them \cite{kaur}. The highly ``usable'' nature of the application programming interface (API), which allows data scientists to quickly produce visualizations without having to understand how things work under the hood, encourages this combination of high trust in results and poor understanding of methods. This relates to our account in Section \ref{values} of how black-box predictive modeling culture arose out of Breiman's government contracting milieu, where obtaining results was prioritized over understanding. Ironically, this push for obtaining results has been extended to the case of explanation methods, which ostensibly exist in order to improve understanding.

\section{Towards pluralistic, contextual, and interactive approaches to explanation}
\label{constructive}

Based on examples of overly universal approaches to ML explanations, and on the values espoused by feminist epistemologists, we suggest some preliminary steps towards creating explanations that are more consonant with feminist epistemic values. Much of the following advice is not necessarily specific to feature importance methods: many parts are also applicable to other aspects of ML. In part, this reflects the fact that ML as a field has not paid much attention to feminist epistemology, so that there are ample opportunities to apply feminist epistemological methods to almost any aspect of ML.

\subsection{Incorporating subjugated points of view}

In line with the idea that the viewpoints of the marginalized may have a crucial critical edge over the perspectives of the powerful, we should place more emphasis on ways for subjugated populations to participate in designing ML explanations. Participation was identified as a crucial quality for feminist design in HCI by Shaowen Bardzell, who argued that since knowers are not substitutable for one another, ``ongoing participation and dialogue among designers and users can lead to valuable insights that could not be achieved scientifically'' \cite{bardzell2010feminist}. Participatory machine learning practices \cite{martin2020participatory,wikipedia} could be extended from participating in algorithm design to participating in the design of explanations, albeit with an awareness that participatory design is not a cure-all \cite{sloane2020participation}.

Incorporating the perspectives of marginalized populations is particularly important when they are stakeholders in the decisions being explained, for example, if they are the subjects of automatic allocations of resources like health or housing. Given the demographics of ML practitioners \cite{ainow2017}, the designers of algorithms making such decisions are unlikely to have a range of social experiences similar to the decision subjects'. Leaving design decisions about end-user explanations up to data scientists would thus likely exclude subjugated points of view.

\subsection{Evaluating explanations contextually}
\label{context}

One important recommendation in feminist epistemology is to acknowledge the situatedness of knowledge---that different knowers have different perspectives. A situated investigation, then, is one that ``forefronts the details of the context of specific people and places at particular points in time, rather than trying to study a system or question with an abstract approach removed from social context'' \cite{katell2020toward}. Elish and boyd \cite{elish2018situating} argue that the act of developing machine learning models itself should be framed as a situated practice, since humans are engaged in personally interpreting data during the development process. So, too, should any attempt to incorporate explanations into a machine learning system.

In evaluating explanations in a feminist spirit, we should consider various contextual factors that affect how well an explanation can perform its social function, such as how explanations are interpreted by stakeholders, unintended consequences, how explanations perform in realistic applications as opposed to toy datasets or problems, and what contexts exist that could cause an explanation method to fail. By bringing in contextual considerations, feminist epistemology enlarges the problem definition of traditional analytic epistemology to include social and ethical considerations. Similarly, one feminist direction that ML explanations could expand in is enlarging traditional problem definitions for work on explanations. Currently, a large proportion of work follows this template \cite{ig,shap, lrp, shrikumar2017learning}:
\begin{enumerate}
    \item Propose a new method of generating explanations for black-box models.
    \item Show that the new method satisfies certain desirable properties. Often, these properties are defined by axioms and the method is proven to satisfy them. Sometimes, the method is shown to satisfy them by carrying out computational experiments. Rarely are the allegedly desirable properties validated by considering what uptake they'll get in different social contexts and different audiences.
    \item  Apply the method to one or more datasets. Use it to explain predictions trained and made on the dataset(s). Validate that these explanations ``make sense.''
\end{enumerate}

(3) Is typically the last step of evaluation, but occasionally, researchers include an additional step of conducting user studies on the explanation method. Test subjects are often recruited via general surveying sites like Amazon's Mechanical Turk, and asked to evaluate explanations with little to no context about why the predictions are being made and what purposes the explanations would serve in a real-world applications \cite{ribeiro2016should, shap-tree, slack2019assessing, shen2020useful}. Test subjects are typically asked to evaluate the explanation on relatively context-free properties such as using the provided explanations to improve the algorithm or simulating the model's predictions, without explaining what the predictions are being used for \cite{ribeiro2016should, poursabzi2018manipulating}.

However, it is hard to evaluate the virtues of explanations without considering the context in which they are meant to operate---the audience, the goals of the explanation, the type of model or data they are applied to, etc. An explanation that best helps data scientists improve a model is not necessarily the best type of explanation for a customer whose request has been denied by an algorithm---if, for example, the most important features according to the explanation are ones that the customer cannot change \cite{barocas-counterfactual}. For similar reasons, explanations that are tailored to satisfy legal requirements like GDPR are also not necessarily the most helpful explanations for decision subjects or model builders \cite{edwards2017slave}. Model builders may benefit more from more sophisticated or technical explanations that are harder for general audiences to understand.

To apply lessons from feminist epistemology to ML explanations, a first step would be to evaluate explanations in a more context-sensitive manner. Axioms should be considered in relation to how desirable they are in specific contexts. User studies on different types of consumers of explanations, rather than on generic consumers, would help. In line with the epistemic advantages that may accrue to marginalized persons, user studies focusing on groups marginalized by algorithms may be particularly helpful in surfacing contextual harms that might be overlooked by more privileged groups. Benchmarks or gold standards tend to impose universal, context-insensitive norms, so they should be used with extreme caution or avoided, as we argued in section \ref{universal}. While using benchmarks may foster an impression that the results are more objective, testing explanation methods on users trying to solve real problems, real (and novel) datasets, all embedded in real social contexts, would provide more comprehensive and nuanced information on the practicalities of implementing the methods. De-contextualizing explanation methods allows for easier comparison across contexts, but also abstracts away from social factors that potentially affect the methods' usefulness.

\subsection{Encouraging seamfulness and pluralism}

Designers have proposed the idea of ``seamful'' design as a way to encourage technology users to avoid over-trusting one interpretation and keep multiple interpretations in mind \cite{seamful}. These ideas can also be applied when we design visualizations for ML explanations. As an antidote against data scientists over-trusting explanations \cite{kaur}, creators of explanatory outputs could design for the following effects \cite{sengers2006staying}:

\begin{enumerate}
    \item Purposefully block the most obvious interpretations of any visualizations, especially if they are misleading, in order to stimulate new ways of understanding the visualizations. One example where this would be helpful is in designing visualizations of log-odds probabilities\footnote{SHAP, a popular post-hoc explanation method, displays feature importance values for classification algorithms in a log-odds space \cite{log-odds}.} to thwart the more ``obvious'' way of reading those quantities in a more intuitive probability space. Another possibility is to design feature attribution charts in a way that reminds viewers that the separate ``bars" displayed for each feature aren't actually separate in reality, due to strong interactions and correlations between features.
    \item Downplay the explanation's authority. We should work on ways to indicate how charts may be inaccurate or misleading. For example, perhaps feature attributions for strongly correlated features can be presented in a more ``uncertain'' manner relative to feature attributions for less correlated features. Here, the uncertainty quantification literature could be particularly helpful \cite{uncertainty}. Alternatively, instead of presenting feature importances in the familiar forms of bar charts or a list of numbers, they could be presented in \emph{less familiar} ways, providing room for reinterpretation of what they mean and thwarting tendencies to understand them in the same way one understands other bar charts. In an example described in \cite{sengers2006staying}, the frequencies of various email categories were represented by changes in a plant's shape. By presenting the data in an unfamiliar form, the artists invite active interpretation and reflection, subverting naively straightforward interpretations.
    \item Design explanatory visualizations in a way that is ambiguous and thwarts any one authoritative interpretation. One concrete suggestion for doing so is to integrate the spatial distortions that dimensionality reduction algorithms produce directly into cluster visualizations \cite{benjamin2019materializing}. While this might seem to be at odds with the highly quantitative nature of most explanation methods, the presence of numbers attached to particular features doesn't mean that those numbers are accurate or adequate representations of the features' actual importances, especially when the explanation method is tailored to satisfy axioms that don't hold in the real world (see Sections \ref{universal} and \ref{data-vs-model}). Given the uncertain relationships between those numbers and the actual features in the data, visualizing them as though they are certain and have unambiguous importance values is misleading. For example, one can imagine an interface that includes multiple explanatory accounts of a model and helps users see the differences between them. In contrast, we currently have multiple, discrete explanation methods that each present their own seemingly authoritative accounts, hiding the uncertainty that is inherent in each of them.
\end{enumerate}

\subsection{Acknowledging the role of problem definition}

Section \ref{context} outlined a ``recipe" for writing a paper about a new ML explanation method. That this ``recipe" exists suggests some kind of common understanding about what the ``problem" of ML explanation is, and what counts as a solution to it. One of the lessons of feminist epistemology is to question how boundaries are drawn around objects. The recipe described suggests a certain limited view of what an explanation is, which largely leaves out social context except at the possible but infrequent step of conducting a user study. If we instead question the boundaries of the problem, new research projects open up.

The field of human-computer interaction (HCI) has helped in expanding the boundaries of ML explanation research by doing better user studies \cite{kaur} and exploring the social functions of ML explanations \cite{hohman2019gamut, hong2020human}. In other parts of ML, HCI has also added new perspectives by studying how improving user interfaces can improve ML systems even without intervening on the algorithm \cite{schnabel2018improving}. Others have studied how co-creating models with end users leads to more more explainable rules and more generalizable models \cite{yang2019study}.

The increased contributions to this field from HCI is a positive trend that will hopefully accelerate the incorporation of feminist epistemic values into ML explanations. One can only hope for more work that integrates traditional technical work with human-centered approaches, as advocated in \cite{human-centered}.

\subsection{Using formalisms critically}

As explained in Section \ref{feminism}, feminist epistemology has long taken a critical stance towards fully formalized systems, instead emphasizing the interactive nature of knowledge creation and the importance of exploring multiple possible meanings. Other researchers have remarked on the tendency of ML to over-value formal proofs \cite{lipton2018troubling}, and the doubts we raise above about axiomatic approaches to explanation can be seen as following this line of thought.

However, jettisoning formalisms altogether is not a solution either, since explaining a formally defined ML algorithm requires engaging with formalisms at some level. Rather, a feminist approach would use formalisms when appropriate, with a careful attention to what possibilities a particular formal framework excludes or renders unsayable, and what it inadvertently favors as prescriptive \cite{agre1997computation}. When one does use a formal framework, its limitations and idealizations should be clearly acknowledged.

\subsection{Using interactive approaches to explanation}

As explained in Section \ref{feminism}, feminist epistemology emphasizes interactional ways of knowing, where the object of study and the knower are engaged in a back-and-forth conversation, rather than a one-directional process where the knower decodes a passive object. Most mainstream ML explanation methods require minimal interaction between the user and the algorithm before creating the explanation, conforming more to the latter mold of an active knower and a passive object. The user typically enters the features they are interested in and the type of explanation they want (e.g. local or global), and gets a static visualization or some numbers in return.

HCI research suggests there might be more interactive ways of creating explanations. One example is Gamut, a visual analytics system that provides an interactive interface to support the interpretation of generalized additive models \cite{hohman2019gamut}. Users found Gamut to be helpful and wanted to use it to understand their own data. One possible way in which interactivity could be extended beyond what Gamut did is by having users interactively design explanation methods, rather than just having an interactive interface for generating pre-determined types of explanations.

\section{Conclusion}

We've attempted to apply feminist epistemological values to ML feature importance methods, diagnosing how the ways we construct and evaluate these methods may insufficiently incorporate these values. Many explanation methods over-emphasize theoretical properties that may seem desirable in idealized contexts but have questionable usefulness in many applicational contexts. Popular feature importance methods also embed strong instrumentalist values that arose from a government contracting and defense milieu. Practitioners value computational modularity in the form of plug-and-play software packages for generating explanations. All of these tendencies---de-emphasizing real-world data, instrumentalism, and computational modularity---are value-laden and should not be considered to be objectively desirable in all contexts.

In particular, we have argued that popular methods of constructing and evaluating feature importance methods do not sufficiently incorporate feminist epistemological values such as context-sensitivity, critically interrogating boundaries, taking the expertise of subjugated knowers seriously, pluralism, and interactional ways of knowing. This is not to say that these popular methods are useless, but to point to the uniformity of values that they embody, and suggest further possibilities that fall outside the existing paradigm. To this end, we've suggested some directions in which future research on explanations can proceed, guided by lessons from feminist epistemology. In our view, we can incorporate feminist values by better incorporating subjugated points of view into the process of designing explanations, evaluating explanations more contextually, resisting the temptation to create benchmarks, thinking critically about how problem definitions favor some values over others, using formalisms with a critical eye, and working on interactive approaches to constructing explanations.

Our proposals mirror recent calls to move towards more human-centered approaches to machine learning \cite{human-centered}. To this end, it's heartening that user studies on ML explanations have started appearing in major HCI conferences \cite{kaur, hong2020human}. However, at mainstream, ``technical'' ML conferences, this kind of work is still largely relegated to special workshops rather than being in the main program. As ML expands its reach into human life, we hope that ML researchers will also correspondingly acknowledge that this broader reach necessitates allowing a broader set of methodologies into its boundaries and including more diverse approaches towards evaluating explanations.

\begin{acks}
We are grateful to the FAccT reviewers and Suresh Venkatasubramanian for their feedback and suggestions, as well as to Miriah Meyer and the members of her Research Paradigms for Human-Centered Computing seminar for the thoughtful discussions which inspired this work.
This research was partially supported by the NSF under grant EAGER-2041960.
\end{acks}

\bibliographystyle{ACM-Reference-Format}
\bibliography{sample-base}


\end{document}